# Thermal Conductivity of PAAm Hydrogel and Its Crosslinking Effect

*Ni Tang†, Zhan Peng†, Rulei Guo†, Meng An, Xiaobo Li, Nuo Yang\* and Jianfeng Zang\**

N. Tang, Prof. J. Zang
School of Optical and Electronic Information
Huazhong University of Science and Technology
Wuhan 430074, People's Republic of China
E-mail: jfzang@hust.edu.cn

Prof. J. Zang
Innovation Institute
Huazhong University of Science and Technology
Wuhan 430074, People's Republic of China

Z. Peng, R. Guo, M. An, Prof. X. Li, Prof. N. Yang
State Key Laboratory of Coal Combustion
Huazhong University of Science and Technology
Wuhan 430074, People's Republic of China
E-mail: nuo@hust.edu.cn

Z. Peng, R. Guo, M. An, Prof. X. Li, Prof. N. Yang
Nano Interface Center for Energy
School of Energy and Power Engineering
Huazhong University of Science and Technology
Wuhan 430074, People's Republic of China

† These authors contributed equally to this work.

As the interface between human and machine becomes blurred, hydrogel incorporated electronics and devices have emerged to be a new class of flexible/stretchable electronic and ionic devices due to their extraordinary properties, such as soft, mechanically robust and biocompatible. However, heat dissipation in these devices could be a critical issue and remains unexplored. Here, we report the experimental measurements and equilibrium molecular dynamic (EMD) simulations of thermal conduction in polyacrylamide (PAAm) hydrogels at room temperature. The thermal conductivity of the PAAm hydrogels can be modulated from 0.33 to 0.51 Wm$^{-1}$K$^{-1}$ by changing the crosslinking density. The crosslinking density dependent thermal conductivity in hydrogels is explained by the competition between the increased conduction pathways and the enhanced phonon scattering effect. The assumption is further supported by both the equilibrium swelling ratio measurement and molecular simulation of hydrogels. Our study offers fundamental understanding of thermal transport in soft materials and provides design guidance for hydrogel-based devices.

Human bodies are mainly composed of soft hydrogels, such as living tissues and muscles. Hydrogels are regarded as one big molecule on macroscale with cross-linked polymer network containing significant amounts of water (70-99 wt %).[1] These intrinsic half-liquid and half-solid characteristics bring hydrogels broad applications such as tissue engineering,[2] cell encapsulation,[3] drug delivery,[4] and soft actuators.[1] Addition to above traditional applications, the emerging hydrogel-based applications are focused on soft electronics, soft robotics and machines.[5] Most applications involving grasping, movement and swelling, thus numerous studies have been carried out to enhance their mechanical properties, such as stretchability, young's modulus, and fracture energy.[6]

Besides mechanical properties, heat dissipation in hydrogel-based electronic devices, as well as soft electronics, however, remains largely unexplored, which may be arisen from the fact that hydrogels used in above domains usually fulfill simple actions based on environmental sensitivity without much participating of electronic component. While for more sophisticated and integrated applications, some energy consuming electronic components and devices, such as conductive wires, rigid electronic devices, and other functional components need to be embodied into biocompatible, soft while mechanically robust hydrogels.[6d] More inspiring examples include ionic hydrogel-based highly stretchable and transparent touch panel,[7] the steerable smart catheter tip realized by flexible hydrogel actuator,[8] double-networking hydrogel-based optical fibers for strain sensors[9] and so on. The attempt on the integration of soft hydrogels with rigid machines shows prominent potential to revolutionize flexible electronics, soft robotics and even medical technologies. The possible heat dissipation at the hydrogel-machine interface with encapsulated electronics components could be a serious issue and need a better understanding.

Existing thermal studies involving hydrogels mainly focused on their nanocomposites and their thermal stabilities[10] and thermal diffusivity.[11] These investigations pay much attention to the role of nanofillers, but not on pure hydrogels and their thermal transportation properties. The hydrogel nanocomposites incorporated with boron nitride (BN) nanopalettes shows an enhanced thermo-responsive time in poly(N -isopropyl- acrylamide) (PNIPAm)/BN hydrogels or a better performance as thermal interface materials in poly(acrylic acid) (PAA)/BN hydrogels.[12] A compelling experimental investigation of intrinsic thermal conductivity of hydrogels itself has not yet been well explored.

Here, we report the first experimental investigation of thermal conduction in hydrogels using 3-omega method. The polyacrylamide (PAAm) hydrogel is chosen as a model hydrogel since PAAm is popular for soft electronics and ionic conductors. Both the experimental and simulation results show that thermal conductivity have an obvious dependence on crosslinking concentrations, which is regarded as a key factor for hydrogels' mechanical properties.[6e] The resultant thermal conductivities of PAAm are in a range of 0.33±0.06 ~ 0.51±0.03 $Wm^{-1}K^{-1}$, which is slightly below the thermal conductivity of pure water (~0.6 $Wm^{-1}K^{-1}$). In order to explore the physical mechanism, we perform equilibrium swelling ratio measurements and equilibrium molecular dynamic (EMD) simulations. We also demonstrate the intuitionistic heat dissipation of an electrically heated graphite rod in hydrogels using an infrared thermal imaging camera.

As prepared PAAm hydrogels are clear and transparent as shown in **Figure1a**. PAAm are made from acrylamide (AAm) monomers consisting of carbon double bond and the –$CONH_2$ group, which could be co-polymerized by crosslinkers, e.g., N–methylenebisacrylamide (MBAm), as shown in Figure1b and 1c.

Various technologies have been developed to measure thermo-physical properties of materials in different states and conditions, such as laser-flash method, transient plane source method, 3-omega method, and etc. When it comes to hydrogel samples, which is unique for their intrinsic half-liquid and half-solid characteristics, it is challenging and special consideration is needed. For example, laser-flash method is a simple one and has been reported to measure thermal conductivity of hydrogel nanocomposites with BN nanosheets.[12a] But laser-flash method usually requires opaque samples and thus is not appropriate for the completely transparent samples. And we think that thermal conductivity of hydrogels is temperature-dependent, so meaurement including heating process, such as transient plane source method, is not suitable as well. 3ω method is chosen here because 3ω method has shown to be successful in measuring samples in liquid,[23] in addition to bulk[14] and thin film samples.[15]

Figure 1d presents the schematic diagram of our experimental setup of 3-omega method for the thermal conduction measurement of the hydrogels. A platinum wire with a diameter of 20 um connected by four copper electrode rods is immersed in hydrogels. The Pt wire serves as both heater and thermometer. When an AC electrical current at angular frequency $1\omega$ is applied at two electrodes, a small voltage signal across the heater to another two electrodes could be detected. The voltage at a frequency of $3\omega$ carrying the thermal effect signal is selected. The actual experimental setup is shown in **Figure S1**. Combining the relationship among the frequency, voltage and temperature enhancement (**Figure S2**), the thermal properties of hydrogels can be extracted. Thus, this class of measurement is aptly known as "3 omega" method. According to the literature,[16] the thermal conductivity can be deduced from 3ω voltage and frequency as follows:

$$\kappa = \frac{\alpha V_{1\omega}^3}{8\pi\iota R} \frac{1}{dV_{3\omega}/d\ln\omega} \tag{1}$$

where κ is the thermal conductivity of the sample, α is the temperature coefficient of the heater. For Pt wire in our experiment, α is 0.00354 PPm ℃$^{-1}$. $V_{3\omega}$ is the 3ω voltage of the heater, $l$ is the length of the heater, and $R$ is the resistance of the heater before heated. More details on the description of 3-omega method can be found in the Supporting Information.

The feasibility of our measurement setup is validated with test of deionized water at room temperature. The measured thermal conductivity of deionized water is 0.60 Wm$^{-1}$K$^{-1}$, which shows a good agreement with the data in NIST database REFPROP. Multiple measurements were conducted on every hydrogel samples to ensure the reliability and reproducibility of our results. The thermal conductivities of PAAm hydrogels as a function of crosslinker concentrations measured by 3-Omega method are shown in the blue curves in **Figure 2**. At the low crosslinker concentration range of 0.016 to 0.099 mol %, the thermal conductivity of hydrogels increases fast and almost linearly from 0.33±0.06 to 0.51±0.03 Wm$^{-1}$K$^{-1}$. With the crosslinker concentration further increasing to 0.263 mol %, the thermal conductivity decreases to 0.33±0.04 Wm$^{-1}$K$^{-1}$. All the thermal conductivities of hydrogels measured in our experiments are below 0.60 Wm$^{-1}$K$^{-1}$. This means, the thermal conductivity of hydrogels consisting of crosslinked polymer network with significant amount water seems always below that of pure water.

To obtain the effective crosslinking density in PAAm hydrogels and understand the corresponding thermal transport behaviors, we implement equilibrium swelling experiment. The water will be imbibed into the polymer network and stretch the polymer chains when the hydrogels are soaked in water. **Figure 3a** presents the time dependent swelling weight of PAAm hydrogels with different crosslinker concentrations. The weight of the hydrogels increases fast at first and will reach saturation when hydrogels are in their equilibrium state after a few days' swelling. The Flory-Rehner equation to correlate the effective crosslinking density $N$ (mol m$^3$) and equilibrium volumetric swelling ratio $Q$ is given by:

$$v_p = Q^{-1} \tag{2}$$

$$-[\ln(1-v_p) + v_p + \chi v_p^2] = NV_s\left[v_p^{1/3} - \frac{v_p}{2}\right] \tag{3}$$

where $v_p$ equilibrium volume fraction of polymer in the hydrogels, $N$ is the crosslinking density (mol/m$^3$), $V_s$ is the molar volume of the DI water, which is $1.8\times10$-5m$^3$mol$^{-1}$ and $\chi$ is the interaction parameter. Detailed analysis of this measurement is included in supporting information. As shown in Figure 3b, the corresponding calculated effective crosslinking density increases fast and linearly in the crosslinker concentrations range of 0.016 to 0.099 mol%, then increases slowly when the concentration is raised from 0.099 to 0.263 mol%.

In amorphous polymer, previous studies have revealed that the effective crosslinking density will basically affect thermal conductivity in two ways.[17] On one hand, the addition of covalent crosslinking bonds will increase thermal conduction pathways between prior non-bonded chain segments. On the other hand, more crosslinking bonds will introduce more phonon scattering along the backbone chains, which reduces the phonon mean free path. Particularly, when the crosslinking density is low (0~10 mol %), the two effects will cancel each other.[17] For large crosslinking density (10~80 mol %), the increase of thermal conduction pathway dominates, which poses a linear increase in thermal conductivity. In our case, crosslinking density is rather small, when the crosslinking density is below 0.099 mol %, the enhancement of thermal conductivity could be attributed to the transition from weak van der Waals interactions to strong covalently bonding interactions, which multiplies the thermal transport pathways. While crosslinking network grows (crosslinker concentration 0.099~0.263 mol %), the amount of phonon scettering sites along backbone chains increases as well. The two competitive mechanisms lead to the variations of thermal conductivity in PAAm hydrogel, as shown in Figure 2.

In order to quantitatively study the thermal conductivity of covalently crosslinked PAAm, EMD simulations based on Green-Kubo method are performed. The simulation cell and its enlarged view are presented in **Figure 4a** and 4b. Taking the PAAm with crosslinking concentration of 2.459 mol % for example, we obtain the time dependent heat current autocorrelation function (HCACF), as shown in Figure 4c. The thermal conductivity extracted from the HCACF curve is shown in Figure 4d. In the EMD simulations, the valance Force Field （CVFF）potentials used for the interactions with bonding as well as non-bonding are presented in **Table S1**. A similar trend of descending after rising in thermal conductivity is also observed in our EMD simulations as shown as the black curves in Figure 2. In EMD simulation, amorphous PAAm chains crosslinked by MBAm are modeled without considering water effect.

Actually, the sufficient water content in hydrogels may contribute significantly to the thermal conduction behaviors of the half-solid and half-liquid samples, which can be inferred from the fact that our EMD simulated results are smaller than the experimental ones. A recently reported research showed that change in water content can greatly affect thermal conductivity of PAA hydrogel in a range of 0.43 $Wm^{-1}K^{-1}$.[12b] Strong intermolecular hydrogen bonds can serve as "soft grips" to restrict the torsional motion of polymer chains and facilitate phonon transport, leading to enhanced thermal conductivities.[18] Besides, with water permeated into the polymer backbone, polymer chains in hydrogel could be stretched. In polymer system, stretching is an effective method to enlarge thermal conductive property in polymer system because of the accompanied better chain alignment and less defects.[11,19] The thermal conductivity of the drawn nanofibers could be as high as 104 $Wm^{-1}K^{-1}$. By the way, in amorphous polymer, heat diffuses more effectively for longer polymer chain.[20] The behind physical mechanism for the thermal transportation in hydrogels involving crosslinking network and water is complicated requires far more efforts from both experimental and theoretical studies.

Furthermore, we demonstrate the heat dissipation behaviors in the hydrogel with thermal conductivity of 0.51 $Wm^{-1}K^{-1}$ by IR thermal imaging camera. As shown in **Figure 5a** and 5b, an electrically heated graphite rod is plugged through the PAAm hydrogel block to provide a heat source and the thermal camera records the time dependent temperature distribution on the top view of samples. Fig. 5c presents that the heat propagates radically from the heat source to the outer circles, generating a 10 ℃ temperature gradient in a distance of 10 mm within 8 min.

**Experimental Section**

*Preparation of hydrogels*: All the following reagents were purchased from Sinopharm Chemical Reagent Co., Ltd. and used without further modified. Samples were synthesized by standard free radicals copolymerization method[6b]. AAm powders were dissolved in deionized water (12 wt %) and mixed with different molar ratio of MBAm as a cross-linker. Ammonium persulfate (APS) in water (3 wt %) as initiator and N, N, N′, N′-tetramethylethylenediamine (TEMED) as the crosslinking accelerator were added into the above solution in sequence. The gel was then sealed under humid condition for future use.

*Equilibrium swelling ratio measurement*: For the equilibrium swelling measurements, PAAm gels with different crosslinker concentrations were cut into a cylinder shape. Then, we dry these samples to constant weight in a vacuum oven. To obtain the equilibrium swelling ratio, dry samples were soaked in deionized water till they reached maximum weight. The weights of the samples as a function of time are recorded.

*EMD simulation*: Atomistic simulations provide a powerful tool for validation and interpretation of experimental results.[21] We have performed EMD simulations of thermal transport in hydrogels with different crosslinking concentrations. In experimental measurements, the weight ratio of water in hydrogel samples with different crosslinking concentrations keeps unchanged. In simulations, to simplify the case, we ignore water effect in hydrogels and only consider interaction between PAAm chains and MBAm. The crosslinking density is set from 0.0017 to 0.17 mol % at 300 K. All the simulations are carried out utilizing the LAMMPS software package.[22] CVFF[23] was used to calculate bonding and non-bonding interactions. The force-field distribution have accurately predicted thermodynamic properties of components in our system.[24] The details of MD simulations can be found in the supporting information.

**Supporting Information**
Supporting Information is available from the Wiley Online Library or from the author.

**Acknowledgements**
This work was supported by the National Natural Science Foundation of China No. 51572096 (J. Z.) and No. 51576067 (N. Y.), and the National 1000 Talents Program of China tenable in HUST (J. Z.).

**Conflict of Interest**
The authors declare no conflict of interest.

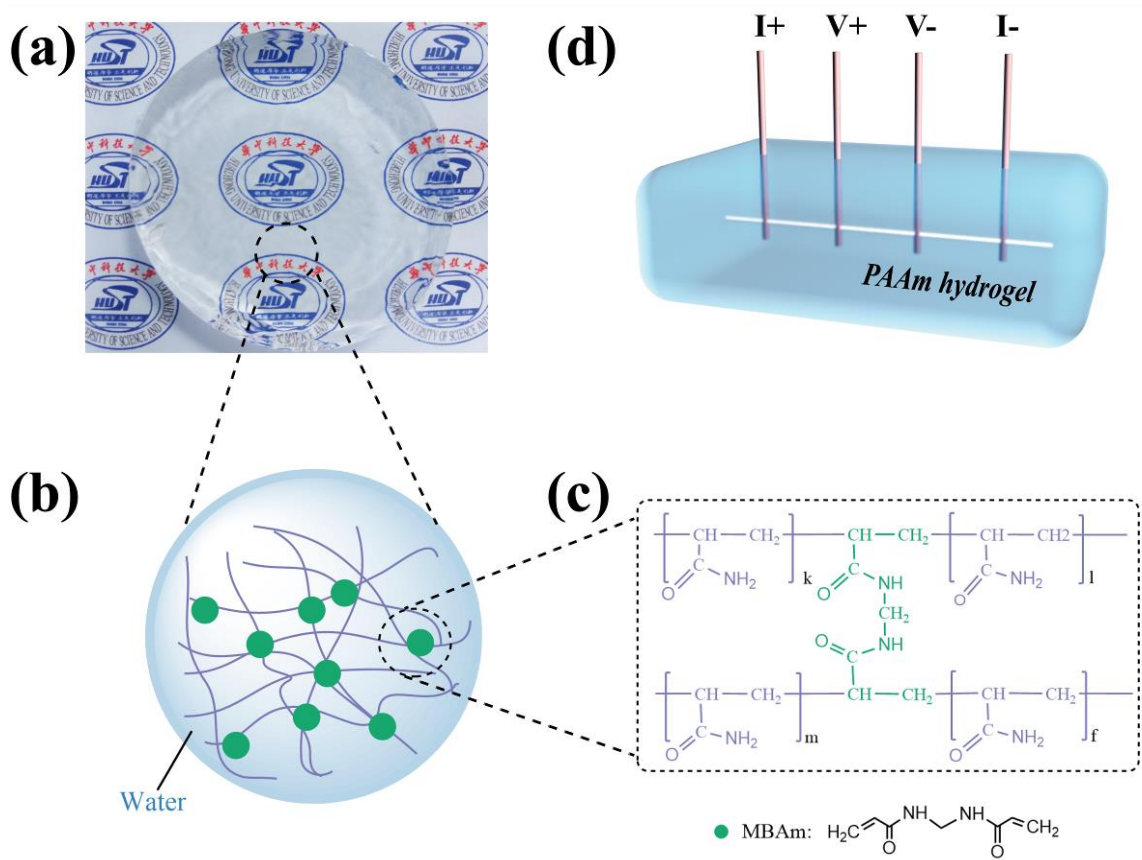

**Figure 1.** PAAm hydrogel and its experimental setup for thermal conductivity measurement. (a) Optical image of the as-prepared PAAm hydrogel sample. (b) Schematic of the polymer network with covalent crosslinking through MBAm (green circles). (c) The molecular structures of the covalently crosslinked polymer chains. (d) The Schematic illustration of 3-Omega method setup for thermal conductivity measurement of hydrogels. A platinum wire is deeply immersed in hydrogels and wired out with four copper probes for applied current and voltage measurements. The four brown rods are copper probes and the white line in hydrogel is the Pt wire.

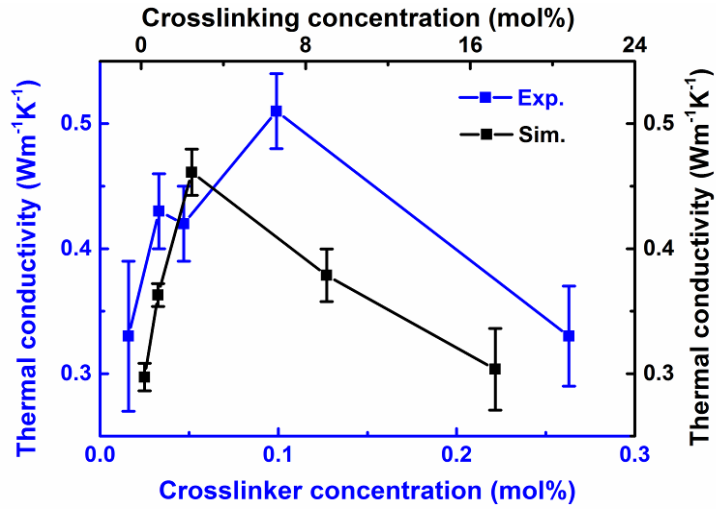

**Figure 2.** Experimental and simulation results on the thermal conductivity of PAAm hydrogels with different crosslinker concentrations. The blue curves are the experimentally measured thermal conductivities by 3-Omega method, and the black curves are the simulated thermal conductivities by molecular dynamics simulations.

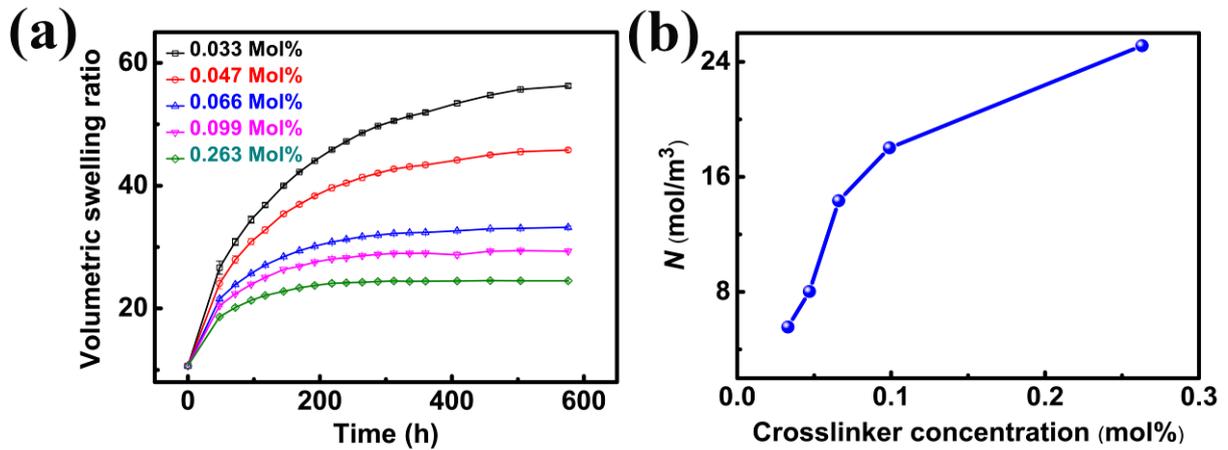

**Figure 3.** Equilibrium swelling measurement for estimation of effective crosslinking density. (a) The volumetric swelling ratio of PAAm hydrogels as a function of swelling time. (b) The relationship of the effective crosslinking density, N, and the corresponding crosslinker concentration.

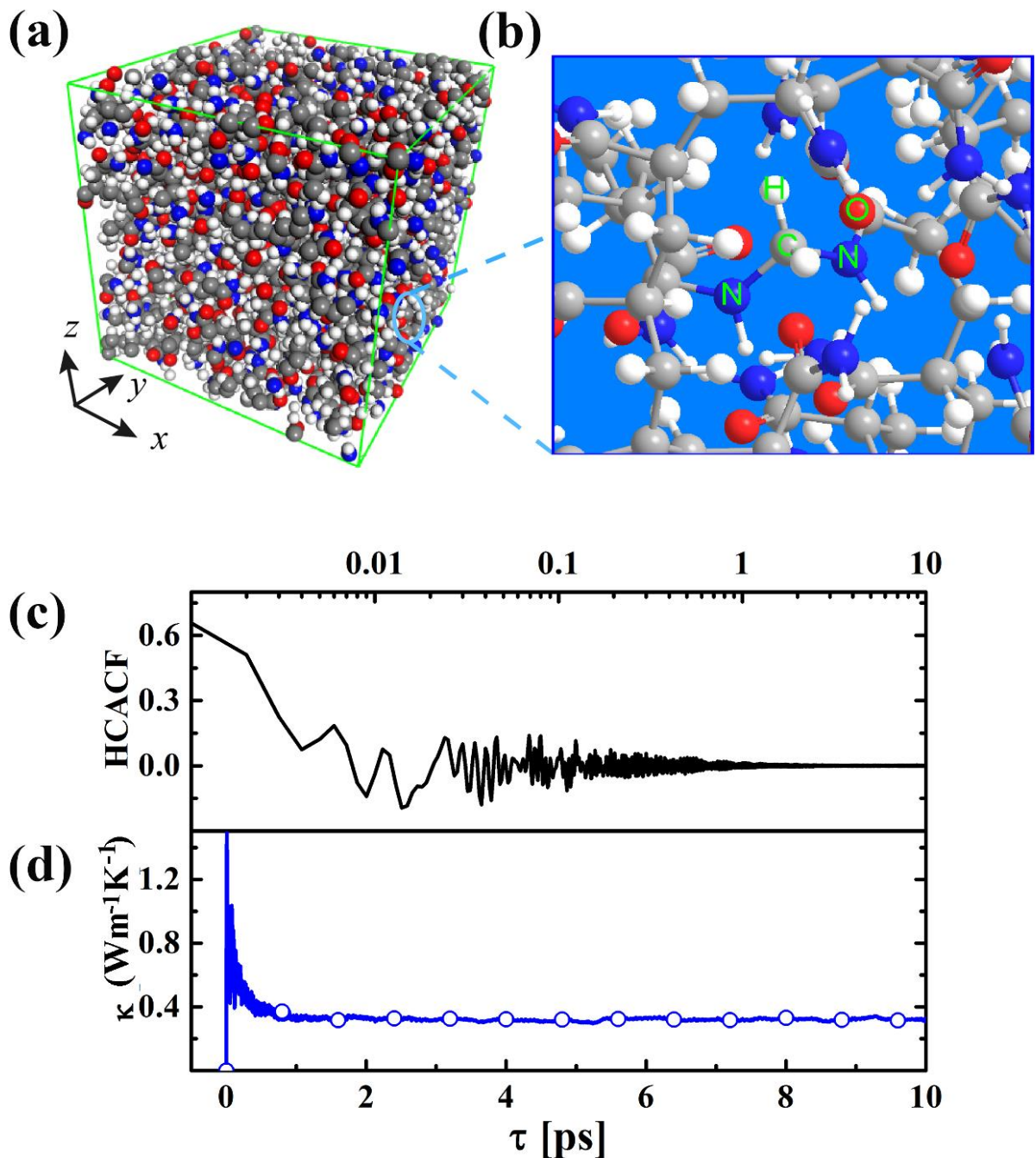

**Figure 4.** Molecular dynamics simulation cell setup and one example of data analysis. (a) Molecular dynamics simulation cell. The simulation cell size is 3.76 nm×3.76 nm×3.76 nm, and its density is 1.36 g/cm$^3$. (b) The enlarged view showing the covalently crosslinked polymer chains. The data analysis of PAAm with 2.459 mol% crosslinking density: (c) time dependent HCACF curve. (d) The lattice thermal conductivity obtained from the time integration of HCACF curve.

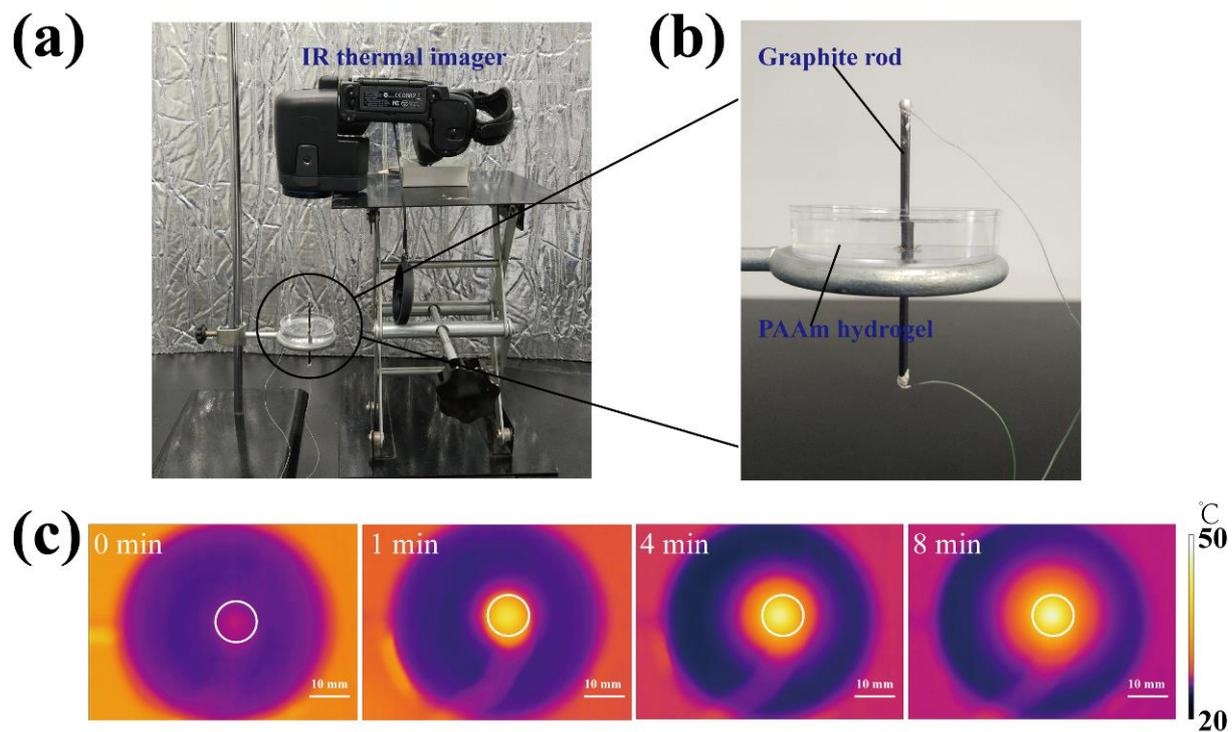

**Figure 5.** A demonstration of heat dissipation behaviors in the hydrogel by IR thermal imaging camera. (a) Optical image of the whole measurement system. The IR thermal camera is fixed on top of the hydrogel sample to record the thermal change. (b) Optical image of the sample. An electrically heated graphite rod is plugged through the PAAm hydrogel sample with thermal conductivity of 0.51 W/m-K. (c) IR thermal images of the hydrogel at different heating time.

Table of Content
**Thermal conductivity of polyacrylamide (PAAm) hydrogels** is measured by 3-omega method. Both experiments and simulation prove that thermal property of PAAm hydrogel can be effectively modulated by its crosslinking density. The mechanism is explained by the competition between the increased conduction pathways and the enhanced phonon scattering effect.

**Keyword**
**hydrogel, thermal conductivity, 3-omega method, crosslinking, molecular dynamics**

Ni Tang, Zhan Peng, Rulei Guo, Meng An, Xiaobo Li, Nuo Yang* and Jianfeng Zang*

**Title**
**Thermal Conductivity of PAAm Hydrogel and Its Crosslinking Effect**

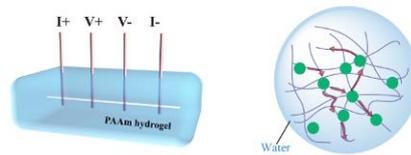

# Supporting Information

**Thermal Conductivity of PAAm Hydrogel and Its Crosslinking Effect**

*Ni Tang, Zhan Peng, Rulei Guo, Meng An, Xiaobo Li, Nuo Yang\* and Jianfeng Zang\**

1. Details of Principle of 3-omega method

A sinusoidal current of frequency ω goes through the platinum wire. Then, the fluctuation of the temperature rise (*ΔT*) is 2ω due to the Joule heating. As the electrical resistance of the platinum wire has a linear relationship with temperature, the fluctuation of the wire resistance is also 2ω. Consequently, the current of frequency ω multiply by the resistance of frequency 2ω produces a small voltage signal of frequency 3ω. This small voltage includes the information of the thermophysical properties of the liquid.

An electrical current of frequency ω and magnitude $I_0$ is applied to the heater.

$$I(t) = I_0 \cos(\omega t) \tag{1}$$

Due to the resistance change of the heater is much smaller than its resistance $R_0$ at environment temperature, the power dissipated by the heater is

$$P = I^2 R \approx I_0^2 R_0 \frac{\cos(2\omega t)+1}{2} \tag{2}$$

For small temperature changes, the resistance of the wire varies linear with temperature as

$$R(t) = R_0[1 + \alpha \Delta T \cos(2\omega t - \Phi)] \tag{3}$$

where $\alpha$ is the temperature coefficient of resistance (TCR),

$$\alpha = \frac{1}{R} \frac{dR}{dT} \tag{4}$$

The resulting voltage across the wire is obtained through the input current multiplying by the heater resistance.

$$V(t) = I(t)R(t) = I_0 R_0 \cos(\omega t) + \frac{1}{2} I_0 R_0 \alpha \Delta T \cos(\omega t - \Phi) + \frac{1}{2} I_0 R_0 \alpha \Delta T \cos(3\omega t - \Phi) \tag{5}$$

In such a structure with an AC electrical current passing through the heater, the heat generated and diffused into the sample can be described by the following equation:

$$\frac{\partial T}{\partial t} = \alpha \left( \frac{\partial^2 T}{\partial r^2} + \frac{1}{r} \frac{\partial T}{\partial r} \right) \tag{6}$$

and the initial and boundary conditions is:

$$r \to \infty, U(r) = 0 \tag{7}$$

$$r = r_0, \frac{\partial T}{\partial r} = -\frac{P}{2\pi r_0 l \kappa} \tag{8}$$

So, solving the equation, we can get

$$\Delta T = \frac{V_{1\omega}^2}{2\pi l R_0 \kappa} \left[ -\frac{1}{2} \ln(\omega) + \frac{1}{2} \ln \frac{D}{r^2} + \ln 2 - r - i \frac{\pi}{4} \right] \tag{9}$$

Where $V_{1\omega}$ the voltage of the heater, and $R_0$ is the resistance of the heater before heated, l is the length of heater, κ is the thermal conductivity of sample, $D$ is the thermal diffusivity of sample, and $r$ is the radius of heater. From Eq. 5, we can get the temperature rise by measuring the 3ω voltage,

$$\Delta T = \frac{2 V_{3\omega}}{V_{1\omega} \alpha} \tag{10}$$

$V_{3\omega}$ is the 3ω voltage of the heater. $\alpha$ is the temperature coefficient of the heater. According to Eq. 9 and 10, we measure the different 3 $\omega$ voltage of different frequency, we can get the thermal conductivity.

2. Experimental Setup

The diagram of the measurement system is shown in **Figure S1**. The platinum wire is welded on copper rods. The diameter of the Pt wire is 20 um. An alternating current (AC) from the signal generator passes though the platinum wire and the resistor. The voltage signal of the wire and the resistor are then input into the Lock-in Amplifier though the differential amplifiers

AMP03. The resistor is adjusted to balance the 1ω voltage signal of the platinum wire because the 1$\omega$ voltage is thousands of times larger than the 3ω voltage.[1] If the 1ω voltage is not eliminated, the obtained 3ω voltage may not be accuracy. The Lock-in Amplifier can get the 3ω voltage signal through the differential input A-B.

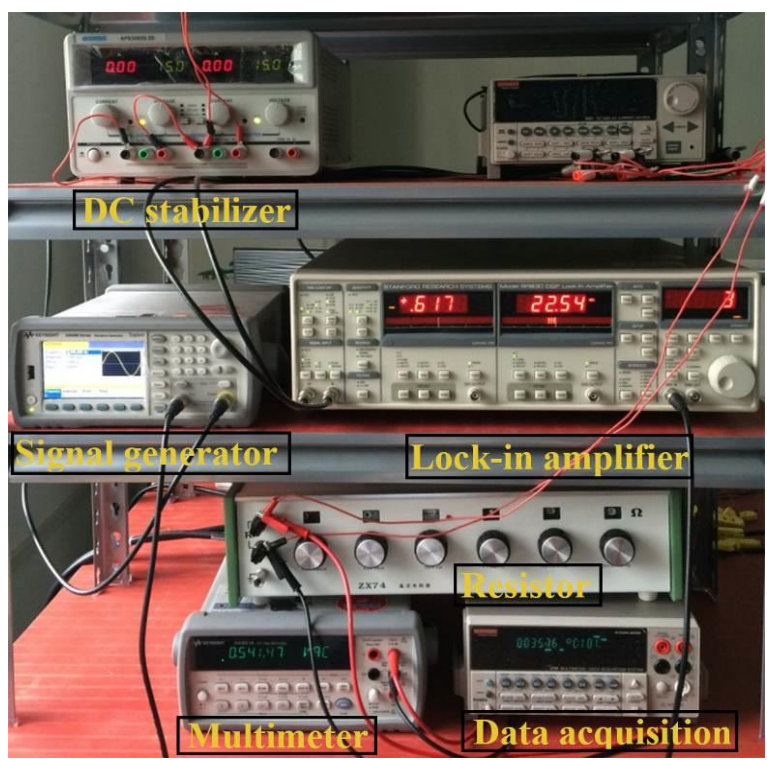

**Figure S1.** (a) Schematic of 3-omega method experimental setup. The red region represents the heating stage for the temperature control. The blue region represent the sample. (b) Picture of 3-ω method experimental setup.

According to the relationship between the length of the wire and the penetration depth, the minimum frequency we can estimate to 0.007 Hz. Similarly , the radius of the wire must be much more smaller than the penetration depth,[2] we can also estimate the maximum frequency to about 7 Hz. So ,the measured frequency should be included in the range from 0.007 Hz to 7 Hz. **Figure S2** exhibits the measured 3ω voltage as a function of Logω for hydrogel with 35 ul concentration at 295.1 K. As the relation of 3ω voltage and Logω is not linear when the frequency is larger than 10 Hz, the lower frequency is only used to calculate. The thermal conductivity is 0.44 Wm$^{-1}$K$^{-1}$. The same procedure is applied to the other samples.

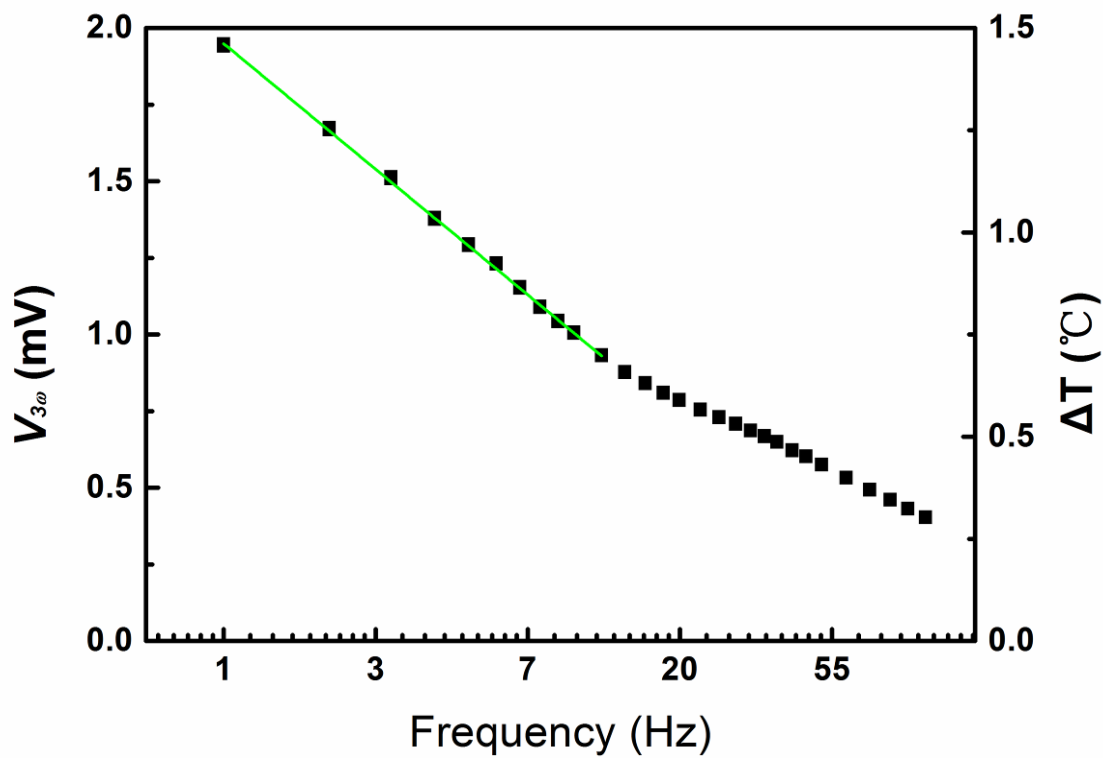

**Figure S2.** Experimental data for sample with 0.033wt% crosslinker concentration at room temperature 295.1K. The black square is the measured value of 3ω voltage. The green solid line is the linear fitting. At low frequency, V$_{3\omega}$ has a linear relationship with logarithmic frequency.

3. Equilibrium Swelling ratio measurement

Phenomenon of limited swelling of hydrogels is a typical characteristic of polymers possessing network structures. Flory and Rehner interpreted the swelling behaviour with entropy theory: As more and more solvent is absorbed (dissolved) by the polymer the network is progressively expanded, there exist two opposite entropies, that is, the entropy of chain configuration and the osmotic or mixing entropy. Equilibrium swelling will be attained when these opposite entropies become equal in magnitude. So we could use the measurement of equilibrium swelling ratio to evaluate the actual crosslinking concentration and network structures of the corresponding PAAm hydrogels. The equilibrium volumetric swelling ratio Q was calculated by equation (11):

$$Q = \frac{V_1+V_2}{V_2} = \frac{\frac{W_1}{\rho_1}+\frac{W_2}{\rho_2}}{\frac{W_2}{\rho_2}} = \frac{\rho_2}{\rho_1} \cdot \frac{W_1}{W_2} \quad (11)$$

where $V_1$ and $V_2$ are the volume of water fraction after equilibrium swelling and gel after drying, $\rho_1$ and $\rho_2$ represent the density of water fraction after equilibrium swelling and the polymer, respectively.

The equilibrium volume fraction of polymer in the hydrogels $v_p$ is:

$$v_p = Q^{-1} \quad (12)$$

Flory-Rehner equation, which is based on the equilibrium swelling of polymer in solvent for three dimensional networks of randomly coiled chains, is suitable for our materials. The Flory-Rehner equation is given as:

$$-[\ln(1-v_p) + v_p + \chi v_p^2] = NV_s\left[v_p^{1/3} - \frac{v_p}{2}\right] \quad (13)$$

where $N$ is the crosslinking density (mol/m$^3$), $V_s$ is the molar volume of the DI water, which is $1.8\times10$-5m$^3$mol$^{-1}$. $\chi$ is the interaction parameter.

The average molecular weight $\overline{M_c}$ between crosslinks is given as:

$$-[\ln(1-v_p) + v_p + \chi v_p^2] = \frac{\rho_2}{M_c}V_s \cdot v_p^{1/3} \quad (14)$$

According to the above expressions, the crosslinking density $N$ increase with increasing of equilibrium swelling ratio, and the average molecular weight of a chain between adjacent crosslinking points shows an opposite trend compared with crosslinking density.

4. MD Simulation details

The lattice thermal conductivity of hydrogels is calculated by equilibrium molecular dynamic(EMD) simulation based on the Green-Kubo formula[3] as

$$\kappa = \frac{1}{3\kappa_B T^2 V} = \int_0^\infty \langle \vec{J}(\tau) \cdot \vec{J}(0) \rangle \, d\tau \tag{15}$$

where κ is the thermal conductivity, $k_B$ is the Boltzmann constant, $V$ and $T$ is the volume of the simulation cell temperature, respectively. $\vec{J}(\tau) \cdot \vec{J}(0)$ is the heat current autocorrelation function (HCACF). The angular bracket denotes ensemble average. The heat current is given by

$$\vec{J}(\tau) = \sum_i \vec{v}_i \, \varepsilon_i + \frac{1}{2} \sum_{i,j} \vec{r}_{ij} (\vec{F}_{ij} \cdot \vec{v}_i) \tag{16}$$

where $\vec{v}_i$ and $\varepsilon_i$ are the velocity vector and energy (kinetic and potential) of particle $i$, respectively. $\vec{r}_{ij}$ and $\vec{F}_{ij}$ are the inter-particle separation vector and force vector between particles $i$ and $j$, respectively. In MD simulation, the temperature $T_{MD}$ is calculated from the kinetic energy of atoms according to the Boltzmann distribution:[4]

$$\langle E \rangle = \sum_{i=1}^{N} \frac{1}{2} m v_i^2 = \frac{1}{2} N \kappa_B T_{MD} \tag{17}$$

All the simulations are carried out utilizing the LAMMPS software package. Consistent Valance Force Field（CVFF）potential was used for bonded as well as non-bonded interactions. The potential details is shown in **Table S1**. The force-field have successfully predicted accurate thermodynamic properties of interests for our system of interest.[5]

The velocity Verlet algorithm is employed to integrate equation of motion, and the time step is set as 0.1fs. At the beginning of a simulation for a thermal conductivity prediction, the system is run in the NVT ensemble to set the temperature. After 50 ps, the simulations run in the NPT ensemble for 500 ps to relax the structure. After relaxation, the converged values of both the cell size and the potential energy obtained, which make sure that there is no stress or strain effects. Then the simulation is switched to run in the NVE ensemble, and the HCACF is obtained over 500 ps. The thermal conductivity is then obtained from the integral of the HCACF. (**Equation 15**).

**Table S1** Force field parameters adopted in CVFF for molecular dynamics simulations.

| Bond Parameters | | |
|---|---|---|
| bond | $K_r$ (eV/Å$^2$) | $r_{eq}$ (Å) |
| C2–H | 14.770 | 1.105 |
| C2–C1 | 13.994 | 1.526 |
| C1–H | 14.770 | 1.105 |
| C1-C | 12.276 | 1.520 |
| C-O | 26.682 | 1.230 |
| C-N2 | 16.825 | 1.320 |
| N2-H | 19.837 | 1.026 |
| C=-C= | 28.411 | 1.330 |
| C=-H | 15.680 | 1.090 |
| C2-C= | 13.998 | 1.500 |
| C-C= | 13.998 | 1.500 |
| C-N | 16.825 | 1.320 |
| C2-N | 16.373 | 1.460 |
| N-H | 20.964 | 1.026 |

| Angle Parameters | | |
|---|---|---|
| Angle | $K_\theta$ (eV/Å$^2$) | $\theta_{eq}$ (degrees) |
| H-C2-H | 1.712 | 106.4 |
| C1-C2-H | 1.925 | 110.0 |
| C1-C2-C1 | 2.021 | 110.5 |
| C2-C1-H | 1.925 | 110.0 |
| C2-C1-C2 | 2.021 | 110.5 |
| C2-C1-C | 2.021 | 110.5 |
| C-C1-H | 1.951 | 109.5 |
| C1-C-O | 2.949 | 120.0 |
| C1-C-N2 | 2.320 | 114.1 |
| N2-C-O | 2.949 | 120.0 |
| C-N2-H | 1.626 | 115.0 |
| H-N2-H | 1.431 | 125.0 |
| C=-C=-H | 1.466 | 121.2 |
| C2-C=-C= | 1.570 | 122.3 |
| C-C=-C= | 1.570 | 122.3 |
| C2-C=-C | 2.168 | 120.0 |
| C=-C2-H | 1.925 | 110.0 |
| O-C-C= | 2.168 | 120.0 |
| N2-C-C= | 2.320 | 114.1 |
| C1-C-N | 2.320 | 114.1 |
| O-C-N | 2.949 | 120.0 |
| C2-N-C | 4.813 | 118.0 |
| C-N-H | 1.626 | 115.0 |
| C2-N-H | 1.518 | 122.0 |
| N-C2-H | 2.233 | 109.5 |

| Van der Waals Parameters | | |
|---|---|---|
| atom type | $\varepsilon$ (eV) | $\sigma$ (Å) |
| C2 | 0.0017 | 3.8754 |
| C1 | 0.0017 | 3.8754 |
| C | 0.0064 | 3.6170 |
| N2 | 0.0073 | 3.5013 |
| O | 0.0099 | 2.8598 |
| C= | 0.0064 | 3.6170 |
| N | 0.0073 | 3.5012 |
| H | 0.0017 | 2.4450 |

| Dihedral Parameters | | | |
|---|---|---|---|
| Dihedral type | K (eV) | d | n |
| H-C2-C1-H | 0.00686 | 1 | 3 |
| H-C2-C1-C2 | 0.00686 | 1 | 3 |
| H-C2-C1-C | 0.00686 | 1 | 3 |
| C1-C2-C1-H | 0.00686 | 1 | 3 |
| C1-C2-C1-C2 | 0.00686 | 1 | 3 |
| C1-C2-C1-C | 0.00686 | 1 | 3 |
| C1-C-N2-H | 0.0651 | -1 | 2 |
| O-C-N2-H | 0.0651 | -1 | 2 |
| C=-C2-C1-C2 | 0.00686 | 1 | 3 |
| C=-C2-C1-H | 0.00686 | 1 | 3 |
| C=-C2-C1-C | 0.00686 | 1 | 3 |
| C2-C=-C=-H | 0.177 | -1 | 2 |
| C-C=-C=-H | 0.177 | -1 | 2 |
| C1-C2-C=-C= | 0.00915 | 1 | 3 |
| H-C2-C=-C= | 0.00915 | 1 | 3 |
| C1-C2-C=-C | 0.00915 | 1 | 3 |
| H-C2-C=-C | 0.00915 | 1 | 3 |
| O-C-C=-C= | 0.0195 | -1 | 2 |
| N2-C-C=-C= | 0.0195 | -1 | 2 |
| O-C-C=-C2 | 0.0195 | -1 | 2 |
| N2-C-C=-C2 | 0.0195 | -1 | 2 |
| N2-C-C=-C2 | 0.0195 | -1 | 2 |
| C=-C-N2-H | 0.0650 | -1 | 2 |
| C1-C-N-C2 | 0.139 | -1 | 2 |
| C1-C-N-H | 0.0520 | -1 | 2 |
| O-C-N-C2 | 0.165 | -1 | 2 |
| O-C-N-H | 0.0781 | -1 | 2 |